\documentstyle[12pt]{article}
\def\figcap{\section*{Figure Captions\markboth
        {FIGURECAPTIONS}{FIGURECAPTIONS}}\list
        {Figure \arabic{enumi}:\hfill}{\settowidth\labelwidth{Figure
999:}
        \leftmargin\labelwidth
        \advance\leftmargin\labelsep\usecounter{enumi}}}
 \relax
\def\np#1#2#3{Nucl.\ Phys.\ B#1 (19#3) #2}
\def\pl#1#2#3{Phys.\ Lett.\ #1B (19#3) #2}
\def\pr#1#2#3{Phys.\ Rev.\ D #1 (19#3) #2}
\def\prb#1#2#3{Phys.\ Rev.\ B #1 (19#3) #2}
\def\prep#1#2#3{Phys.\ Rep.\ #1 (19#3) #2}

\def\rmp#1#2#3{Rev.\ Mod.\ Phys.\ #1 (19#3) #2}
\newcounter{hran}
\def\bmini#1{\setcounter{hran}{\value{equation}}
\refstepcounter{hran} \setcounter{equation}{#1}
\renewcommand{\theequation}{\thehran\alph{equation}}
              \begin{eqnarray}  }
\def\bminiG#1{
          \setcounter{hran}{\value{equation}}
          \refstepcounter{hran}
          \setcounter{equation}{-1}
          \renewcommand{\theequation}{\thehran\alph{equation}}
          \refstepcounter{equation}
    \label{#1}
          \begin{eqnarray}          }
\def\emini{\end{eqnarray}\setcounter{equation}{\value{hran}}
\renewcommand{\theequation}{\arabic{equation}}}
\newskip\humongous \humongous=0pt plus 1000pt minus 1000pt
\def\caja{\mathsurround=0pt} \def\eqalign#1{\,\vcenter{\openup1\jot
\caja   \ialign{\strut \hfil$\displaystyle{##}$&$
\displaystyle{{}##}$\hfil\crcr#1\crcr}}\,} \newif\ifdtup

\def\re#1{(\ref{#1})}
\def\half{\mbox{\small $\frac{1}{2}$}}

\def\frac#1#2{ {{#1} \over {#2} }}
\def\ie{\hbox{\it i.e.}{ }}      
\def\eg{\hbox{\it e.g.}{ }}      
\def\beq{\begin{equation}}
\def\eeq{\end{equation}}
\def\beeq{\begin{eqnarray}}
\def\eeeq{\end{eqnarray}}
\def\bit{\begin{itemize}}
\def\eit{\end{itemize}}
\def\ben{\begin{enumerate}}
\def\een{\end{enumerate}}

\def\bc{\bar c}
\def\D{\Delta}
\def\d{\delta}
\def\G{\Gamma}
\def\L{ \Lambda}

\def\R{{\mbox{\scriptsize R}}}
\def\UV{$\L_0\to\infty\;$}
\def\BRS{ {\mbox{\footnotesize{BRS}}}}
\def\cut{ {\mbox{\footnotesize{cutoff}}}}
\def\lat{ {\,\mbox{\footnotesize{lat}}}}
\def\A{{\cal {A}}}
\def\K{K_{\L_0}}
\def\O{{\cal O}}
\def\g0{ g_{\L_0}}
\def\gam{\gamma}
\def\aU#1{\L_0^{-#1}\;}
\begin{document}
\begin{titlepage}
\renewcommand{\thefootnote}{\fnsymbol{footnote}}
\begin{flushright}
     UPRF 96-484\\
     IFUM-536-FT\\
     August 1996 \\
\end{flushright}
\par \vskip 10mm
\begin{center}
{\Large \bf
Gauge invariant action at the ultraviolet cutoff
\footnote{Research supported in part by MURST, Italy
and by EC Programme ``Human Capital and Mobility",
contract CHRX-CT93-0357 (DG 12 COMA).}}
\end{center}
\par \vskip 2mm
\begin{center}
        {\bf M.\ Bonini} \\
        Dipartimento di Fisica, Universit\`a di Parma and\\
        INFN, Gruppo Collegato di Parma, Italy\\
        and\\
        {\bf G.\ Marchesini}\\
        Dipartimento di Fisica, Universit\`a di Milano and\\
        INFN, Sezione di Milano
\end{center}
\par \vskip 2mm
\begin{center} {\large \bf Abstract} \end{center}
\begin{quote}
We show that it is possible to formulate a gauge theory starting
from a local action  at the ultraviolet (UV) momentum cutoff which is
BRS invariant.
One has to require that fields in the UV action and the fields
in the effective action are not the same but related by a local
field transformation. The few relevant parameters involved in this
transformation (six for the $SU(2)$ gauge theory), are perturbatively
fixed by the gauge symmetry.
\end{quote}
\end{titlepage}

\vskip 0.6cm
\noindent
Consider the UV (or bare) action $S_{\L_0}[\phi,\gam]$
of a Yang-Mills theory with $\L_0$ the UV cutoff
($\phi$ represents the gauge, ghost and antighost
fields while $\gam$ are the BRS sources \cite{BRS}).
Within the exact Wilson renormalization group (RG) formulation
\cite{W}-\cite{B} this UV action can be viewed as the result of
taking an elementary underlying theory and integrating all degrees
of freedom with frequencies larger than the UV cutoff.
For $\L_0$ much larger than any physical momentum,
$S_{\L_0}[\phi,\gam]$ can be taken local and given by
few ``relevant'' parameters\footnote{
In a functional, the coefficients of local field monomials with non-negative
dimension will be called relevant parameters \cite{W}.}
which are the only trace of the underlying theory.

By using the exact Wilson RG it has been shown \cite{B,YM} that it
is possible to compute perturbatively the relevant parameters of
$S_{\L_0}[\phi,\gam]$ in such a way that the effective action
$\G[\phi,\gam]$, obtained by integrating over the fields with
frequencies smaller than $\L_0$, satisfies the Slavnov-Taylor
(ST) identities in the limit $\L_0\to\infty$.
This is obtained by properly fixing the relevant parameters of 
$\G[\phi,\gam]$ and then computing the
large $\L_0$ dependence of the relevant parameters of the UV action.

Although it is beautiful to see that by fixing the few parameters
in $S_{\L_0}[\phi,\gam]$ one obtains the functional identity
for $\G[\phi,\gam]$ which is
the consequence of the gauge symmetry, it is however ugly the fact
that the UV action itself does not manifest the gauge symmetry
in a simple way. For example in $S_{\L_0}[\phi,\gam]$ one needs to
include an appropriate mass term for the gauge field.
The UV action may be considered as the closest functional to the
underlying theory, but it does not have any explicit symmetry.

In this paper we show that it is actually possible to formulate the
theory in such a way that both the UV action is BRS invariant and
the effective action satisfies the ST identities.
To obtain this we try to adapt to the continuous case the formulation
of lattice gauge theory (LGT) \cite{LGT} in which the UV action and
the path integral measure are explicitly gauge invariant.
They are expressed in terms of link variables given by exponential
of the gauge fields on the lattice.
As we shall see the key point we exploit and adapt to the continuous
case is the non linear relation between the link and field variables.

\vskip .4 cm \noindent
{\bf UV action and BRS symmetry}
\vskip .2 cm \noindent
We start by introducing the UV action in the continuum
and, as an example, we consider the $SU(2)$ Yang-Mills theory.
Inspired by LGT we are ready to distinguish between the fields
$\phi=(A_\mu,c,\bc)$ which enter into the physical observables,
\ie in the effective action $\G[\phi,\gam]$, and the UV fields
$\Phi=(\A_\mu,c,\bc)$, which enter into the UV action\footnote{
It is not necessary to introduce different fields for the ghost and 
the antighost. See later.}, the
analogous of the link variables in LGT.
At the UV scale $\L_0$ we assume that the UV action is given by
the BRS classical action, \eg in the Feynman gauge,
\beq\label{BRS}
S_\BRS[\Phi,\gam ;\g0 ]
=\int d^4x \biggl\{
-\frac 1 4 F_{\mu\nu}^{2}
-\frac 1 2 (\partial_\mu \A_\mu)^2
+ W_\mu \cdot D_\mu c
- \frac 1 2 \, v \cdot c  \wedge c
\biggr\}\,,
\eeq
with
$\;F_{\mu\nu}=\partial_\mu \A_\nu- \partial_\nu \A_\mu +
\g0  \A _\mu \wedge \A _\nu$,
$\;D_\mu c  =\partial_\mu c  +\g0 \A_\mu \wedge c $,
$\;W_\mu=\partial_\mu \bc +u_\mu/\g0$,
and the UV coupling $\g0 $.
We have introduced the usual scalar and external $SU(2)$
products and the sources $\gam=(u_\mu,\,v)$ for the BRS variations of
$\A_\mu$, and $c$ respectively.

In the path integral over the field $\Phi(x)$ we have to consider only
frequencies smaller than $\L_0$.
To cutoff larger frequencies one could for instance add
\cite{P}-\cite{YM} to the UV action the following contribution (in
terms of the Fourier transform of the fields)
\beq\label{S2}
S_\cut[\Phi]
\equiv
- \int \frac{d^4p}{(2\pi)^4}\;p^2\;(\K^{-1}(p)-1)\;\left\{
\half \A^\R_\mu(-p)\cdot\A^\R_\mu(p)- c^\R(-p)\cdot c^\R(p)
\right\}
\,,
\eeq
in which we have introduced the cutoff function
$\K(p)=1$ for $p^2 < \L_0^2$ and rapidly vanishing for
larger frequencies.
The contribution $S_\cut[\Phi]$ vanishes if the fields are in the low
momentum region ($p^2<\L_0^2$) while diverges if the fields have
large frequencies ($p^2 > \L_0^2$). As a consequence, in the path
integral the large frequency fields are suppressed.
$S_\cut[\Phi]$ is not BRS invariant, however its variation
always involves large frequency fields which are suppressed in
the path integral.

For conveniency (see \cite{YM}) we have introduced in \re{S2} the
``renormalized'' fields $\Phi^\R$ related to the
UV fields $\Phi$ by multiplicative constants to be discussed later.

\vskip .4 cm \noindent
{\bf Ultraviolet and effective fields}
\vskip .2 cm \noindent
We have now to establish the relation between the ``UV field''
$\A_\mu(x)$ entering into the UV action, and the ``effective field''
$A_\mu(x)$ entering into the effective action $\G[\phi,\gam]$.
In general one should assume that the two fields are the same
within a distance $1/\L_0$.
The UV field $\A_\mu(x)$ is then given by a functional of the
effective field $A_\mu(y)$ with $y$ confined in a region
$|x-y|<1/\L_0$. By expanding around $x$ one finds
\beq\label{1}
\A_\mu(x)=A_\mu(x)
\left\{ 1-\aU 2 \D_2(A(x))-\aU 4\D_4(A(x)) +\cdots \right\}
\,,
\eeq
where the dots denote terms with higher powers of $1/\L_0$.
The functions $\D_2(A)$ and $\D_4(A)$ of dimension two and four
respectively are polynomials in the fields $A_\nu$ which are
scalars under the Lorentz and $SU(2)$ transformations.
This requirement fixes the polynomials in terms of six parameters
\beq\label{D24}
\eqalign{
\D_2(A)=
&
\half c_5 A_\nu\cdot A_\nu \,,
\cr
\D_4(A)=
&
 \half c_0 A_\nu \cdot \partial^2 A_\nu
-\half c_1 (\partial_\nu A_\nu)^2
+c_2 (\partial_\nu A_\rho) \cdot A_\rho \wedge A_\nu
\cr&
+\frac 14 c_3 (A_\rho\wedge A_\nu)\cdot(A_\nu\wedge A_\rho)
+\frac 14 c_4 (A_\nu\cdot A_\nu)^2
\,.
}
\eeq
Formally, in the limit \UV the UV and effective fields coincide.
However this is not the case and the limit is not trivial, as we shall see.

In LGT, the UV field $\A_\mu(x)$ corresponds to the link variable
$U_\mu(x)$  and the field transformation \re{1} corresponds to the
exponential mapping $U_\mu(x)=\exp \{ g_\lat\;aA_\mu(x)\}$
with $g_\lat$ the lattice gauge coupling and $a$ the lattice size.
This transformation has the advantage that it induces a group
structure, thus one can easily prove that the lattice effective action
satisfies ST identities \cite{BRSlat}. However it cannot be applied
in a simple way to the continuous case since in the exponential
mapping Lorentz covariance is lost.
In the continuous case the transformation \re{1} preserves
Lorentz covariance and the expansion parameters $c_i$ will be fixed by
requiring ST identities for the effective action $\G[\phi,\gam]$.

Consider the BRS transformation of the elementary field $\Phi$
which leaves invariant the UV action \re{BRS} and the measure $D\Phi$. 
For the gauge field one has
\beq\label{BRSel}
\d_\BRS\; \A_\mu(x)=\eta
\{ \partial_\mu c  +\g0 \A_\mu(x) \wedge c(x)\}
\,,
\eeq
with $\eta$ the Grassmann parameter.
For the effective field this transformation becomes
\beq\label{BRSef}
\d_\BRS\; A_\mu(x)=\left\{ \d^{(0)}_\BRS +\aU 2\d^{(2)}_\BRS
                +\aU 4\d^{(4)}_\BRS +\cdots \right\}A_\mu(x)
\,,
\eeq
where
\beq
\eqalign{
& \d^{(0)}_\BRS A_\mu(x)= \eta
\{ \partial_\mu c  +\g0 A_\mu(x) \wedge c(x)\}
\cr&
\d^{(2)}_\BRS A_\mu(x)= \eta
c_5 \{ \partial_\mu c(x) \,(\half A_\nu(x)\cdot A_\nu(x))
+A_\mu(x)\, (\partial_\nu c(x)\cdot A_\nu(x))
\}
\,,
}
\eeq
and a more complex expression for $\d^{(4)}_\BRS A_\mu(x)$.
Formally the BRS transformation of $A_\mu(x)$ and $\A_\mu(x)$
are the same in the limit \UV. However, due to the non linearity
of \re{1}, the Jacobian of the transformation is not trivial even
for \UV. To compute the Jacobian consider
\beq\label{dF}
\frac{\d \A^a_\mu(x)}{\d A^b_\nu(y)}
= \d^4(x-y)
\left\{  \d_{ab}\d_{\mu\nu}
 -\aU 2 c_5 \;
   [ \d_{ab}\d_{\mu\nu} \half A_\rho\cdot A_\rho + A^a_\mu \; A^b_\nu]
 -\aU 4  \frac{\partial (A_\mu^a\;\D_4(A))}{\partial A_\nu^b}
 +\cdots
\right\}
\,,
\eeq
where the fields in the curly bracket are at the point $x$.
The logarithm of the Jacobian is
\beq\label{Jac}
\mbox{Tr}\;\ln\frac{\d \A}{\d A}
=-S_m[\phi] + \O(\aU 2)
\,,
\eeq
where $S_m[\phi]$ is the finite contribution for \UV
\beq\label{Sm}
\eqalign{
S_m[\phi] =
&
\int d^4x \; \{
  7 c_5 \L_0^2 A_\nu \cdot A_\nu
+ 7 c_0 A_\nu \cdot \partial^2 A_\nu
- 7 c_1 (\partial_\nu A_\nu)^2
+15 c_2 (\partial_\nu A_\rho) \cdot A_\rho \wedge A_\nu
\cr&
+ 4 c_3 (A_\rho\wedge A_\nu)\cdot(A_\nu\wedge A_\rho)
+ (4c_4+\frac52 c_5^2)\, (A_\nu\cdot A_\nu)^2
\}
\,.
}
\eeq
Therefore, although \re{1} formally is trivial for \UV\,
the logarithm of the Jacobian does not vanish in this limit
but contains six relevant terms.
Among these parameters one has a gauge mass term proportional
to the UV scale.

\vskip .4 cm \noindent
{\bf Effective action and ST identities}
\vskip .2 cm \noindent
We now consider the effective action $\G[\phi,\gam]$ obtained
by adding to the UV action the source term expressed in terms of the
effective fields $\phi$
\beq\label{source}
(j,\;\phi)=\int d^4x \left\{ j_\mu(x)\cdot A^\R_\mu(x)
+\bar\chi(x)\cdot c^\R(x) +\bc^\R(x)\cdot \chi(x)
\right\}
\,.
\eeq
We introduce the renormalized fields and BRS sources $\phi^\R,\gam^\R$
related to $\phi,\gam $ by normalization constants
\beq\label{wfn}
A_\mu=\sqrt{Z_A}\; A^\R_\mu
\,, \;\;\;\;
c    =\sqrt{Z_c}\; c^\R
\,, \;\;\;\;
W_\mu=\sqrt{Z_c}\; W^\R_\mu
\,, \;\;\;\;
v    =\sqrt{Z_v}\; v^\R
\,.
\eeq
The effective action $\G[\phi,\gam]$ is obtained as the
Legendre transform
\footnote{
In the path integral the source $j$ are coupled to renormalized
field $\phi^\R$ as in  \re{source}.
However, for simplicity of notation, we denote by $\phi$ the
fields in the effective action, \ie the ``classical'' fields
conjugate to the source $j$ in the Legendre transform of $Z[j,\gam]$.
Similarly, in the effective action we denote by $\gam$ the BRS source.}
of the functional
\beq\label{ZPhi}
Z[j,\gam]=\int D\Phi \exp
\left\{
-S_\BRS[\Phi ,\gam ,\g0] -S_\cut[\Phi] +(j,\;\phi)
\right\}
\,.
\eeq
By making the field transformation \re{1} and using the Jacobian
\re{Jac} we can write
\beq\label{Zphi}
Z[j,\gam]=\int D\phi \; e^{ -S_{\L_0}[\phi,\gam] +(j,\;\phi)}
\,,
\eeq
where $S_{\L_0}[\phi,\gam]$ is the UV action in terms of the
effective fields
\beq\label{bare}
S_{\L_0}[\phi,\gam]
=S_\BRS[\phi ,\gam ,\g0] + S_m[\phi] +S_\cut[\phi] + \cdots
\eeq
where the dots represent irrelevant contributions proportional to
powers of $\aU 2$.
The relevant part $S_{\L_0}[\phi,\gam]$ depends on nine parameters:
the six constants $c_i$ in the field transformation \re{1},
the UV coupling $\g0$ and the three normalization constants of
the field and BRS sources \re{wfn}.
Notice that $c_0$ has not to be considered as an independent parameter
since it can be absorbed in $Z_A$.

While the UV action $S_\BRS[\Phi,\gam,\g0]$ and the measure $D\Phi$
are both invariant under the BRS transformation \re{BRSel},
the UV action $S_{\L_0}[\phi,\gam]$ in \re{bare} and the measure
$D\phi$ are not invariant under the BRS transformation \re{BRSef}.
This is due to the fact that the logarithm of the Jacobian from $\Phi$
to $\phi$ has been included into $S_{\L_0}[\phi,\gam]$.

We come now to the question whether the effective action obtained
from \re{Zphi} does satisfy the ST identities in the limit 
$\L_0\to\infty$.
As mentioned at the beginning the requirement that $\G[\phi,\gam]$
satisfy the ST identities fixes (perturbatively) the six parameters
$c_i$ of the field transformation \re{1}. Moreover all polynomials
in \re{1} with dimension higher than four can be neglected.
These properties are consequences of the well known results
on renormalizability and ST identities in gauge theories
(see for instance \cite{B} and therein references).

The mentioned results have been recently rederived \cite{B},
\cite{YM} within the framework of exact RG in which one explicitly
deals with a momentum cutoff and a UV action given by the relevant
part of $S_{\L_0}[\phi,\gam]$.
One shows perturbatively that, by properly fixing the nine relevant
parameters in $\G[\phi,\gam]$, the effective action remains
finite for \UV (renormalizability) and satisfies ST identities.
Among these parameters one has the physical coupling $g$ at a
subtraction point $\mu\ne0$.

By using the RG equations one can then compute in terms of $g$
and $\mu/\L_0$ the nine relevant parameter in the UV action
$S_{\L_0}[\phi,\gam]$, obviously the same number of relevant
parameters in $\G[\phi,\gam]$.
In particular one computes the six UV constants of the field
transformation \re{1}. For the explicit values at one loop
see \cite{YM}. At one loop one finds that all $c_i$ with $i\ge 1$
are finite for \UV. Recall that $c_0$ contributes to the
normalization of the gauge field.

\vskip .4 cm \noindent
{\bf Summary and comparison with LGT}
\vskip .2 cm \noindent
By following the example of LGT we have shown that it is possible in
the continuum to formulate a gauge theory in which both the effective
action satisfies ST identities and the ``bare'' action at the UV
cutoff is BRS invariant.
The key point is the assumption that the field in the UV
action are different from the fields in the effective action.
The main difference between the continuous and the lattice formulation
is in the structure of the field transformation.
It is useful to summarize here these differences and the
consequences.

\noindent {\it Continuous case.}
The field transformation \re{1} is constructed in such a way
to satisfy the Lorentz covariant structure of the field and, as a
consequence, one has manifest Lorentz invariance in the UV action
$S_{\L_0}[\phi,\gam]$.
The field transformation involves infinite number of parameters but
one needs to fix only the six coefficients of the field polynomials
\re{D24} of dimension two and four which give rise to relevant
contributions to $S_{\L_0}[\phi,\gam]$.
Polynomials in \re{1} of higher dimensions gives irrelevant
contributions to $S_{\L_0}[\phi,\gam]$.
The coefficients $c_i$ are fixed by the gauge symmetry, \ie by
properly fixing these coefficients one has that the effective
action satisfies the ST identities.

\noindent {\it Lattice gauge theory.}
The field transformation is given by the exponential mapping
$U_\mu(x)=\exp \{g_\lat\;a A_\mu(x)\}$. By using the group property
one can easily prove that the lattice effective action
satisfies ST identities \cite{BRSlat}.
However, the transformation has manifest cubic covariance instead
of the Lorentz one. From the logarithm of the Jacobian one finds
that the UV action, written in terms of the gauge fields $A_\mu(x)$,
contains relevant terms which are non-scalar under Lorentz
transformations. It contains also a gauge field mass term and
infinite irrelevant couplings.
While in the continuous case all irrelevant contributions in the UV
action can be neglected, in the LGT case they are essential to cancel
the various Lorentz non-scalar contributions.

\vspace{3mm}\noindent
We have benefited greatly from discussions with Giuseppe Burgio,
Marco D'Attanasio, Claudio Destri and Enrico Onofri.


\begin{thebibliography}{99}
\bibitem{BRS}
        C. Becchi, A. Rouet and R. Stora, \pl{52}{344}{74}.
\bibitem{W}
        K.G. Wilson, \prb{4}{3174, 3184}{71};
        K.G. Wilson and J.G. Kogut, \prep{12}{75}{74}.
\bibitem{P}
        J. Polchinski, \np{231}{269}{84}.
\bibitem{G}
        G. Gallavotti, \rmp{57}{471}{85}.
\bibitem{B}
        C. Becchi, On the construction of renormalized quantum field
        theory using renormalization group techniques,
        in {\it Elementary particles, Field theory and Statistical
        mechanics}, Eds. M. Bonini, G. Marchesini and E. Onofri,
        Parma University 1993.
\bibitem{YM}
        M. Bonini, M. D'Attanasio and G. Marchesini,
        \pl{346}{87}{95};\np{437}{163}{95}.
\bibitem{LGT}
        K.G. Wilson, \pr{10}{2445}{74}. See also {\it Lattice gauge
        theories and Monte Carlo simulations}, ed C. Rebbi,
        World Scientific 1983 and, 
        for a recent presentation, I. Montvay and G. Munster,
        {\it Quantum fields on a lattice}, Cambridge University press.
1994. \bibitem{BRSlat}
        H. Kawai, R. Nakayama and K. Seo, \np{189}{49}{81};
        T. Reisz, \np{318}{417}{89};
        S. Caracciolo, P. Menotti and A. Pelissetto,
        \np{375}{195}{92}.
\end{thebibliography}
\end{document}